# Fast Spatio-temporal Compression of Dynamic 3D Meshes

Gerasimos Arvanitis

*Electrical and Computer Engineering*

*University of Patras*

Patra, Greece

arvanitis@ece.upatras.gr

Aris S. Lalos

*Industrial Systems Institute*

*Athena Research Center*

Patra, Greece

lalos@isi.gr

Konstantinos Moustakas

*Electrical and Computer Engineering*

*University of Patras*

Patra, Greece

moustakas@upatras.gr

*Abstract*—3D representations of highly deformable 3D models, such as dynamic 3D meshes, have recently become very popular due to their wide applicability in various domains. This trend inevitably leads to a demand for storage and transmission of voluminous data sets, making the need for the design of a robust and reliable compression scheme a necessity. In this work, we present an approach for dynamic 3D mesh compression, that effectively exploits the spatio-temporal coherence of animated sequences, achieving low compression ratios without noticeably affecting the visual quality of the animation. We show that, on contrary to mainstream approaches that either exploit spatial (e.g., spectral coding) or temporal redundancies (e.g., PCA-based method), the proposed scheme, achieves increased efficiency, by projecting the differential coordinates sequence to the subspace of the covariance of the point trajectories. An extensive evaluation study, using different dynamic 3D models, highlights the benefits of the proposed approach in terms of both execution time and reconstruction quality, providing extremely low bit-per-vertexper-frame (bpvf) rates.

## I. INTRODUCTION

The rapid technological advances in real-time 3D scanning and the development of tools and methods for the accurate reconstruction of the 3D models have opened up a variety of new applications in a wide range of domains (like education, entertainment, industry, medicine). Moreover, the easiness of capturing real 3D objects and the creation of synthetic models result in the generation of a huge amount of data that are continuously increasing. This trend stresses the need for the development of novel techniques on dynamic 3D mesh compression, which will be capable to address the aforementioned growing demands.

Throughout the years, several approaches have been proposed to achieve reliable compression results by improving some characteristics of the compression process, like the reconstruction quality, the low computational complexity, and the compression rate. The first attempts for 3D mesh compression occurred by utilizing some traditional compression schemes [1], [2] such as the direct quantization of the 3D coordinates of the vertices. However, this type of quantization introduces significantly high geometrical error to the reconstructed surface of the model and at the same time, the compression rates remain high, requiring a large number of bits-per-vertex for efficient encoding. Other works [3] suggest quantizing the differential or delta coordinates. These approaches succeed in capturing the local relation of vertices and usually outperform the methods that directly quantize the 3D coordinates, especially in small bpvf ranges.

The aforementioned effect is attributed to the fact that quantization errors that may affect the low-frequency components are not perceptually significant to the human's visual system. At the same time, high-frequency components may be removed since high frequencies correspond to noise or smallscale features that cannot be easily distinguished [3]. Building on this line of thought, the authors in [4], [5] suggested performing compression by maintaining only a small number of low-frequency components. The main drawback of the Graph Fourier schemes is the increased demands for computational power since they require the decomposition of large matrices. Orthogonal iterations [6] have been proposed to overcome this limitation providing very fast and accurate results.

In [7], the authors proposed a 3D wavelet transform domain compression method for coding geometry video data to exploit the spatial and temporal correlation at multiple scales. The work presented in [8] is based on the assumption that 3D models are representable by a sequence of weighted and undirected graphs where the geometry and the color of each model can be considered as graph signals. The authors in [9] proposed a spectral clustering-based dynamic reshaping model. After the lossy compression of spatio-temporal segments through PCA, a spectral clustering of all the PCA elements is computed, introducing also three reshaping schemes of the PCA elements within each cluster. In [10], an initial temporal cut was computed to obtain a small subsequence by detecting the temporal boundary of dynamic behavior. Then, they apply a two-stage vertex clustering on the resulting subsequence to classify the vertices into groups with optimal intra-affinities. In other works [11] [12], it was suggested the partition of the sequence into several clusters with similar poses, and then the decomposition of the meshes in each cluster into primary poses and geometric details using the manifold harmonic bases derived from the extracted keyframe in that cluster. The authors of [13] presented two extensions, based on Edgebreaker and TFAN mesh connectivity coding algorithms, of MPEG-I Video-based Point Cloud Compression (V-PCC) standard to support mesh coding. In this paper, we focus on addressing all the aforementioned drawbacks and on exploiting the benefits of different geometrical and spectral approaches by introducing a novel pipeline for dynamic 3D mesh compression exploiting efficiently the spatial and the temporal redundancies. Our main contributions can be summarized as follows:

- We provide a simple and easily reproducible mathematical framework, that combines both geometrical and spectral attributes. The main idea lies on the observation that, on contrary to mainstream state-of-the-art spectral coding approaches, by projecting the spatio-temporal matrix of the $\delta$ coordinates on the subspace of the covariance of the point trajectories (eigen-trajectories), we can derive superior spatio-temporal compression.
- We take advantage of the spatio-temporal coherence between consecutive differential representations and between the eigen-trajectories and graph Fourier subspaces in order to achieve very low compression rates.
- We estimate the temporal coding dictionaries by fast tracking methods based on orthogonal iterations, resulting in very fast implementation, necessary for challenging cases attributed to large 3D animation datasets and demanding execution time constraints.

The rest of this paper is organized as follows: In Section 2, we present some preliminaries and basic definitions. In Section 3, we discuss in detail each step of the proposed method. Section 4 presents the experimental results in comparison with other methods and in Section 5 we draw the conclusions.

II. PRELIMINARIES

Let us assume the existence of a dynamic 3D mesh $\mathbf{A}$ represented by a sequence of $k$ static meshes $M \in R^{n \times 3}$ so that $\mathbf{A} = [M_1; M_2; \cdots ; M_k]$. Each static 3D mesh consists of $n$ vertices represented as a matrix of vertices $\mathbf{V} = [\mathbf{v}_1; \mathbf{v}_2; \cdots ; \mathbf{v}_n] \in R^{n \times 3}$ in a 3D coordinate space where $\mathbf{v} = [v_x; v_y; v_z] \in R^{3 \times 1}$. Any $j$ vertex $\mathbf{v}_j$ that belongs to the neighborhood of the first-ring area $\Psi_i$ is a neighbor of $\mathbf{v}_i$.

The Laplacian matrix $\mathbf{L} \in R^{n \times n}$ can be defined as:

$$\mathbf{L} = \mathbf{D} - \mathbf{C} \quad (1)$$

where $\mathbf{C} \in R^{n \times n}$ is the binary adjacency matrix with elements:

$$\mathbf{C}_{ij} = \begin{cases} w_{ij} = 1 & \text{if } j \in \Psi_i \\ 0 & \text{otherwise} \end{cases} \quad (2)$$

and $\mathbf{D} = diag\{d_1, d_2, ..., d_n\}$ is a diagonal matrix with $d_i = \sum_{j=1}^{n} \mathbf{C}_{ij}$.

The delta coordinates $\delta \in R^{n \times 3}$ of a mesh are calculated as the difference between each vertex $\mathbf{v}_i$ and its neighbors that belong to the first-ring area $\Psi_i$ [14]:

$$\boldsymbol{\delta}_i = [\delta_{x_i}, \delta_{y_i}, \delta_{z_i}] = \mathbf{v}_i - \frac{1}{|\Psi_i|} \sum_{j \in \Psi_i} \mathbf{v}_j, \quad \forall i = 1, n \quad (3)$$

or

$$\delta = \mathbf{LV} \quad (4)$$

where $|\Psi_i|$ is the number of immediate neighbors of $i$.

A. Spectral Analysis

A matrix $\mathbf{R}$ can be decomposed applied the Singular Value Decomposition (SVD) method:

$$[\mathbf{U} \, \mathbf{\Lambda} \, \mathbf{U}^T] = SVD(\mathbf{R}) \quad (5)$$

where $\mathbf{U} = [\mathbf{u}_1, \mathbf{u}_2, ..., \mathbf{u}_n]$ is the orthonormal matrix of the eigenvectors and $\mathbf{\Lambda} = diag\{\lambda_1, \lambda_2, ..., \lambda_n\}$ is a diagonal matrix with the corresponding eigenvalues.

The Graph Fourier Transform (GFT) is defined as the projection of a matrix $\mathbf{K}$ onto the orthonormal matrix $\mathbf{U}$ of the eigenvectors, according to:

$$\hat{\mathbf{U}}_k = \mathcal{G}(\mathbf{K}) = \mathbf{U}^T \mathbf{K} \quad (6)$$

where $\hat{\mathbf{U}}_k \in R^{n \times 3}$ is a matrix representing the GFT of the matrix $\mathbf{K}$ and $\mathcal{G}(.)$ represents the GFT function.

B. Orthogonal Iterations

The direct implementation of SVD has an extremely high computational complexity, making the implementation to be prohibitive for large matrices. A solution to this drawback is the use of subspace tracking algorithms that rely on the execution of iterative schemes, executed in $n_b$ equal-sized blocks of data [15]. The most widely adopted subspace tracking method is the Orthogonal Iterations (OI) [16], providing very fast and accurate solutions, especially if the given initial subspace (i.e., the estimated matrix of eigenvectors of the previous block) is close enough to the original subspace of interest. The initial subspace $\mathbf{U}[1]$ of the $1^{st}$ block has to be orthonormal in order to preserve orthonormality. For this reason, $\mathbf{U}[1]$ is estimated by a direct SVD implementation, while all the following subspaces $\mathbf{U}[i]$, $i = 2, ..., n_b$ are adaptively estimated by the following Algorithm 1, using QR decomposition.

---

Algorithm 1: OI updating process for any $\mathbf{R}[i]$ block of data

```
1  U[1] ← SVD(R[1]);
2  for i ← 2 to n_b do
3    U[i]^(1) ← U[i − 1];
4    for t ← 2 to t_max do
5      U[i]^(t) ← QR(R[i]U[i]^(t−1));
6    end
7    U[i] ← U[i]^(t_max);
8  end
```

## III. PROPOSED SPATIO-TEMPORAL COMPRESSION SCHEME

We start by formulating the spatio-temporal matrices $A_x, A_y, A_z \in \mathbb{R}^{k \times n}$, where the matrix $A_x$ consists of the $x$ coordinates of the $n$ vertices of the $k$ frames of the animated mesh. To mention here that the following process is applied three times, one for each coordinate matrix, however, for the sake of simplicity, we will present it here once, only for the $x$ coordinate. Firstly, we estimate the autocorrelation matrix $R_x \in \mathbb{R}^{k \times k}$

$$R_x = A_x A_x^T \quad (7)$$

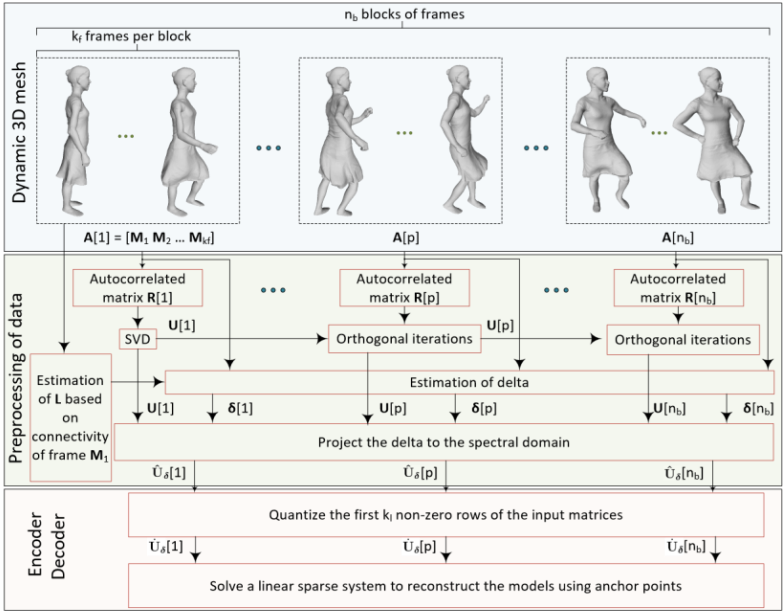

Fig. 1: Pipeline of the proposed method for the compression of dynamic 3D meshes using spatio-temporal information.

and we decompose it, via Eq. (5), in order to estimate the orthonormal matrix $U_x \in \mathbb{R}^{k \times k}$ of the eigenvectors. We also estimate the delta coordinates of the matrix $A_x$:

$$\delta_x = L A_x^T \in \mathbb{R}^{n \times k} \quad (8)$$

We project the delta coordinates to the matrix of the eigenvectors $U_x$, using the GFT function:

$$\hat{U}_{\delta_x} = \mathcal{G}(\delta_x^T) = U_x^T \delta_x^T \in \mathbb{R}^{k \times n} \quad (9)$$

Encoder: The matrix $\hat{U}_{\delta_x}$ represents the information that we want to compress (e.g., for storage, transmission or any other purposes). We take advantage of the observation that the visual representation of matrix $\hat{U}_{\delta_x}$ is very sparse (i.e., a lot of values are close to zero), as shown in Fig. 2. So, we assume that we can encode only the values of its first $k_l$ rows without losing a lot of information, where $k_l < k$. The values of the rest rows are replaced with zeros (or in other words, we encode them using 0 bits).

$$\dot{U}_{\delta_x i} = \begin{cases} Q(\hat{U}_{\delta_x i}), & \text{if row } i < k_l \\ 0 & \text{otherwise} \end{cases} \quad (10)$$

where $Q(.)$ denotes the quantization function. Fig. 3 shows how the selection of the $k_l$ value affects the reconstructed results. Comparing the Figs. 3-(d) and 3-(e), we can conclude that after a current value of $k_l$ (e.g., > 20), maintaining more information does not mean better visual results. Decoder: The decoder has to solve the following Eq. (11) for the reconstruction of matrix $A_x$.

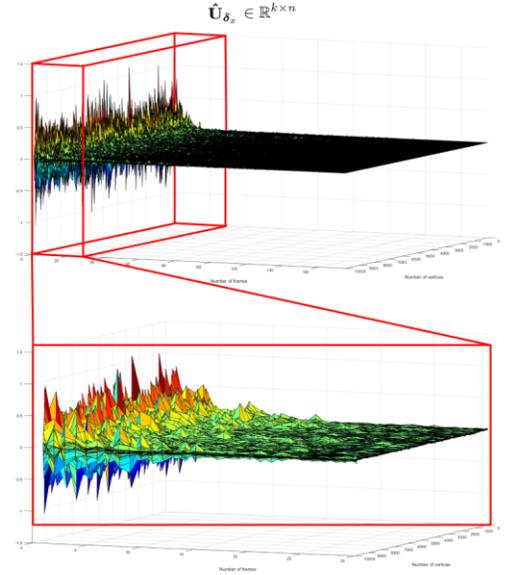

Fig. 2: Visual representation of matrix $\hat{U}_{\delta_x}$, and enlarged detail showing how sparse it is.

$$\tilde{A}_x = L^{-1}(U_x \dot{U}_{\delta_x})^T \in \mathbb{R}^{n \times k} \quad (11)$$

where $\tilde{A}_x$ denotes the reconstructed matrix, $L$ can be estimated by the known connectivity, which is also the same for all frames, and $\mathbf{U}_x$ is a dictionary matrix which is assumed as known.

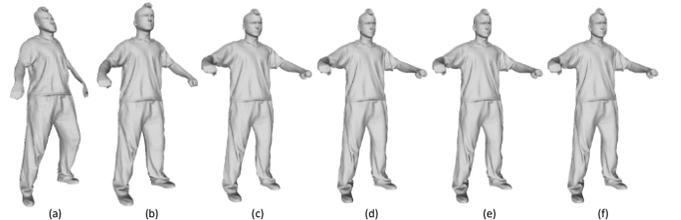

Fig. 3: Reconstructed results of "Handstand" model, by encoding only the first (a) 2, (b) 5, (c) 10, (d) 20, (e) 30 out of 175 rows of the matrix $\hat{\mathbf{U}}_{\delta x}$, (f) original 3D mesh.

Solving the Eq. (11), we take an acceptable solution. However, to further increase the level of detail of the reconstructed 3D meshes, we can also quantize a set of $n_c$ uniformly distributed vertices, known as anchor points, where $n_c$ usually corresponds to the 1% of the total number of vertices $n$. The indices of these vertices represented by the set N. The reconstruction of the 3D mesh vertices is performed as in [3], by solving the following sparse linear system:

$$\begin{bmatrix} \mathbf{L} \\ \mathbf{I}_{n_c} \end{bmatrix} \tilde{\mathbf{A}}_x = \begin{bmatrix} \mathbf{U}_x \dot{\mathbf{U}}_{\delta_x} \\ \mathbf{I}_{n_c} Q(\mathbf{A}_x^T) \end{bmatrix} \quad (12)$$

where $\mathbf{I}_{n_c} \in \mathbb{R}^{n_c \times n}$ is a subset of the identity matrix $\mathbf{I} \in \mathbb{R}^{n \times n}$, since its rows has been constructed by those $i$ rows of $\mathbf{I}$ where $i \in \mathrm{N}$.

### A. Speed-up Process using OI for Online Procedure

To make the process more computationally light, we suggest separating the dynamic 3D mesh into $n_b$ equal-sized blocks of $k_f = k/n_b$ consecutive meshes and then using OI according to subsection II-B. In this case, the dimensions of the matrix $\mathbf{U}_x[1]$, which is the only matrix that is decomposed via SVD, is $k_f \times k_f \ll k \times k$, making the execution of the process to be very fast. Additionally, the reconstruction of the animation in blocks is also faster due to the lower dimensions of the proceeding matrices.

The assumption, concerning the coherence, is based on the observation that blocks of meshes of the same 3D animation maintain both geometric (same connectivity and geometrical features) and temporal characteristics (similar motion between consecutive frames). This coherence leads to a very fast solution since the initialization of the OI starts to a subspace which is very close (i.e., coherent) to the real one. It is worth mentioning here that the OI is a solution that also can be used in real-time applications since both matrix multiplications and QR factorizations have been highly optimized for maximum efficiency on modern serial and parallel architectures.

Nevertheless, despite the fast execution of this approach, the accuracy of the reconstructed results is inferior to these of using the whole sequence of meshes in one block. The advantages of implementing this approach are more apparent when the animated 3D model consists of many frames > 1000 or when the application priority is a fast implementation instead of a very accurate reconstruction. In Fig. 1, the framework of the proposed approach is briefly presented. The case of using the whole animated mesh as input could be assumed as a subcase of this pipeline where $n_b = 1$.

## IV. EXPERIMENTAL ANALYSIS AND RESULTS

For the comparisons, we use different variants that either exploit temporal and/or spatial coherence's. More specifically, we present the results by using: (i) Projection of the delta coordinates to the eigen-trajectories (PCA), (ii) Projection of the quantized delta coordinates to the eigen-trajectories (PCA+q), (iii) PCA and then quantization to the projected vertices (v2v) [6], (iv) (PCA+q) and serial reconstruction using anchor points (PCA+qs), (v) (PCA+q) and parallel reconstruction using anchor points (PCA+qp), (vi) OI into blocks of frames and parallel reconstruction using anchor points (blocks), (vii) per mesh GFT without taking advantage of the temporal information (per mesh) [17].

In Fig. 4, we present the NMSVE metric [4] per each frame for different approaches and models. PCA+qs and PCA+qp approaches provide the best results. We can also see that the NMSVEs, for these methods that exploit the temporal coherence's, follow a similar pattern. This means that the quality of the reconstruction depends on the type of motion of the dynamic model. On the other hand, the "per mesh" approaches provide reconstructed results with almost the same NMSVE value per each frame. Fig. 5 shows the bpv for each row of the matrix $\hat{\mathbf{U}}_\delta$ for different approaches. For most of the methods, only the first $k_l$ rows of the matrix are quantized with bpv > 0. In the case of the "block"-based approach, we need to quantize the first rows of each block separately, increasing however in this way the final bvpf. Additionally, we can observe that in blocks, consisting of similar frames (e.g., the last blocks of these animations where the frames are significantly similar due to a static motion), the required bits in order to achieve the same reconstruction error are less than in other blocks capturing fast motions. In Table I, we present the bpvf for the quantization bits $q$ of the matrix elements $\hat{\mathbf{U}}_\delta$ according to Eq. (10), or based on the average estimation of Fig. 5. We also present the bpvf for the quantization of the anchor points $q_a$, the dictionary matrix $q_d$, as well as the total bpvf $q_s$ which represent the summary of all the quantization bits, as presented in Eq. (13). We assume that for a lossless quantization, both each anchor point and each element of a dictionary matrix is represented using 16 bits.

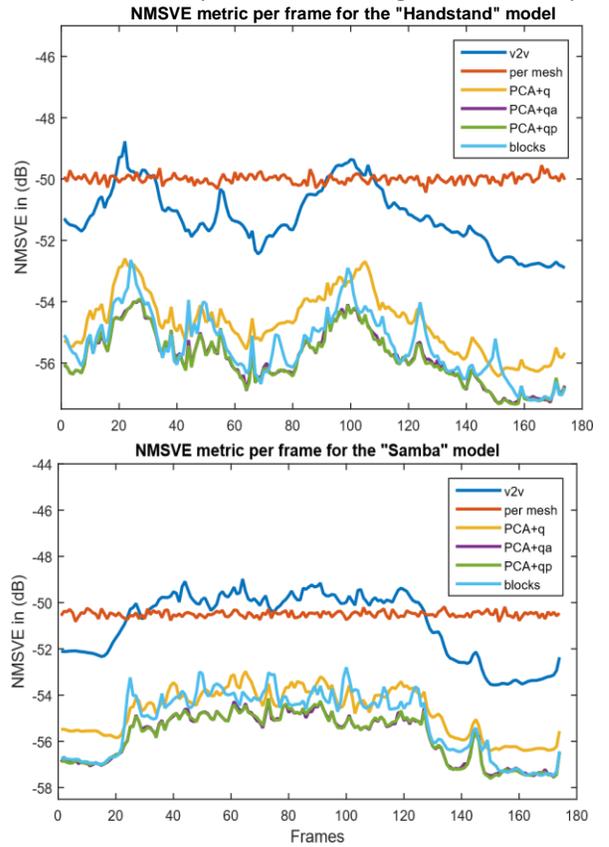

Fig. 4: NMSVE metric per each frame for different approaches and models.

$$q = \frac{bits \cdot n \cdot k}{n \cdot k}, \quad q_a = \frac{bits \cdot n_c \cdot k}{n \cdot k}, \quad q_d = \frac{bits \cdot n_b^2 \cdot k_f \cdot k_f}{n \cdot k} \quad (13)$$
$$q_s = q + q_a + q_d$$

At this point it should be mentioned, that for the "block"-

| Model | Q | PCA | PCA+q | v2v | blocks | PCA+qs PCA+qp |
|---|---|---|---|---|---|---|
| Hand-stand | $q$ | 1.2215 | 0.2893 | 0.5783 | 0.4006 | 0.2893 |
| | $q_a$ | - | - | 0.1600 | 0.1600 | 0.1600 |
| | $q_d$ | 0.2799 | 0.2799 | 0.2799 | 0.1847 | 0.2799 |
| | $q_s$ | 1.5014 | 0.5692 | 1.0182 | 0.7453 | 0.7292 |
| Samba | $q$ | 0.7289 | 0.1827 | 0.4362 | 0.3057 | 0.1827 |
| | $q_a$ | - | - | 0.1595 | 0.1595 | 0.1595 |
| | $q_d$ | 0.2808 | 0.2808 | 0.2808 | 0.1498 | 0.2808 |
| | $q_s$ | 1.0088 | 0.4635 | 0.8761 | 0.6155 | 0.6226 |

TABLE I: Total $q_s$ bpvf for different approaches.

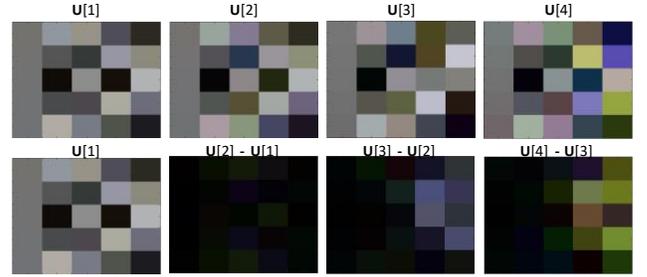

Fig. 6: Consecutive dictionary matrices of handstand model and the corresponding difference matrices.

based approach, we encode the difference U[$i$ + 1] − U[$i$] between two consecutive matrices (second row of Fig. 6), instead of the original dictionary matrices U[$i$], taking advantage of the spatial coherency between the consecutive dictionaries. This information is easier to be quantized using fewer bits.

Table II presents the execution times during the data processing and encoding phase (first row) and the reconstruction phase (second row). The "blocks" approach provides the fastest solution in both cases. For meshes with more than > 20,000 vertices the "per mesh" approach can not be implemented due to the high computational cost for estimating the per frame SVD decomposition.

Fig. 7 depicts the reconstructed results and additional enlarged details of the models for easier visual comparison among the methods. We also provide the STED [21] metric that is ideal for evaluating the quality of the reconstructed ani-mation and a heatmap visualization of the angle $\theta$ representing the difference between surface normals of the reconstructed and the original models. The PCA+qs and the PCA+qp approaches have very similar reconstructed results, as we can see in Figs. 4 and 7, however the PCA+qs method is much slower (Table II) since it needs to solve the Eq. (12) in $k$ times. Additionally, it appears temporal artifacts that are apparent only in the animation mode (non in a static figure).

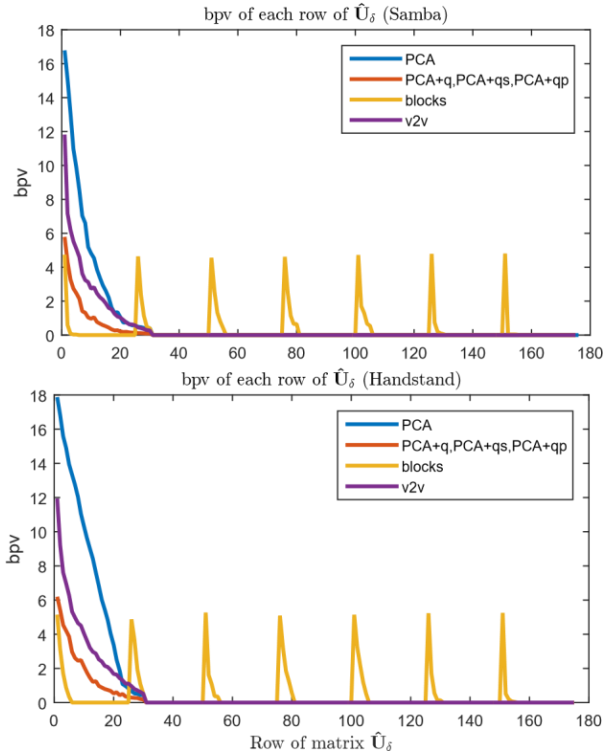

Fig. 5: bpv for different approaches.

| Models (n/k) | Per mesh | PCA +q | PCA +qs | PCA +qp | blocks |
|---|---|---|---|---|---|
| Handstand [18] (10002/175) | 39557.42 | 0.89 | 0.89 | 0.89 | 0.07 |
| | 4736.25 | 26.47 | 4736.25 | 28.67 | 22.12 |
| Dinosaur [19] (20218/152) | - | 0.69 | 0.69 | 0.69 | 0.06 |
| | - | 40.63 | 6920.56 | 45.53 | 37.19 |
| Samba [18] (9971/175) | 37314.55 | 0.87 | 0.87 | 0.87 | 0.07 |
| | 4214.55 | 24.86 | 4214.55 | 26.17 | 20.89 |
| | 22.49 | 0.17 | 0.17 | 0.10 | 0.07 |

| Model | | | | | |
|---|---|---|---|---|---|
| Chinchilla [19] (4307/84) | - | - | - | - | - |
| | 476.04 | 4.41 | 476.04 | 4.68 | 4.22 |
| Camel-g. [20] (21887/48) | - | 0.12 | 0.12 | 0.12 | 0.06 |
| | - | 54.27 | 2730.24 | 56.88 | 51.27 |
| Flag (2704/1000) | 7062.21 | 2.38 | 2.38 | 2.38 | 0.07 |
| | 2633.36 | 2.22 | 2633.36 | 2.37 | 1.84 |

TABLE II: Execution times for the compression and reconstruction of different models.

## V. Conclusions and Future Work

In this work, we presented a pipeline for the efficient compression of dynamic 3D meshes. The method formulates a spatio-temporal matrix and takes advantage of the geometric and temporal coherences that consecutive differential frame coordinates of the same animation have, by projecting the spatio-temporal matrix of the $\delta$ coordinates on the eigentrajectories, we can derive superior spatio-temporal compression. In the near future, we will investigate the ideal number of rows that are needed in order to achieve the optimal results and additionally the best-selected number of blocks that increase the spatio-temporal coherences on consecutive frames. Finally, although it seems that the random selection of anchor points does not significantly affect the final results, we will investigate whether there is an ideal combination of vertices that can provide the best results.

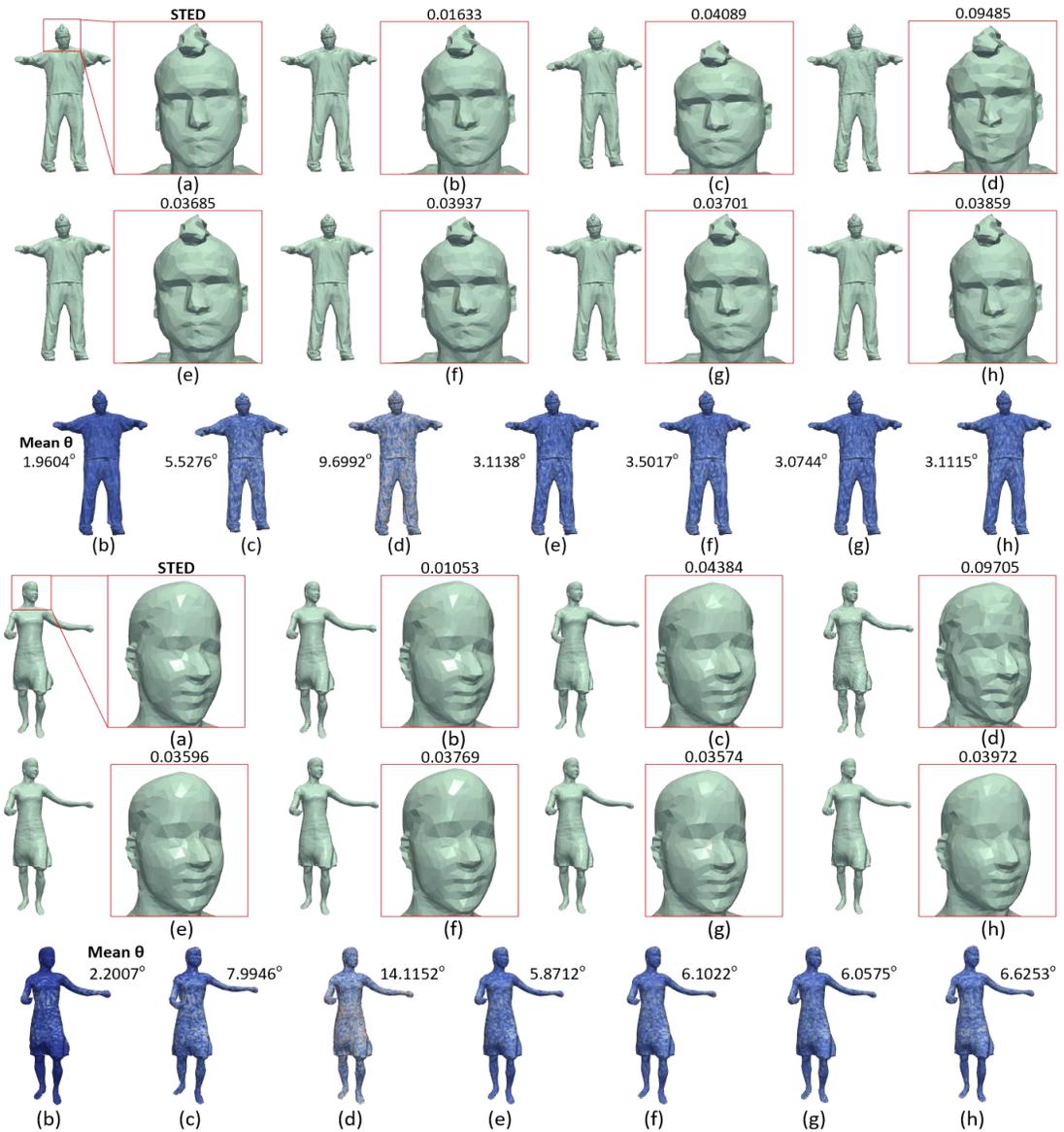

Fig. 7: (a) Original models and reconstructed in (b) PCA, (c) PCA+q, (d) v2v, (e) PCA+qs, (f) PCA+qp using 0.5% of anchor points, (g) PCA+qp using 1% of anchor points, (h) "blocks" approach using 1% of anchor points.